\documentclass[lettersize,journal]{IEEEtran}
\usepackage{array}
\usepackage[caption=false,font=normalsize,labelfont=sf,textfont=sf]{subfig}
\usepackage{stfloats}
\usepackage{url}
\usepackage{verbatim}
\usepackage[utf8]{inputenc}
\usepackage{cite}
\usepackage{amsmath,amssymb,amsfonts}
\usepackage{amsfonts,latexsym}
\usepackage{amsthm}
\usepackage{multirow}
\usepackage{graphicx}
\usepackage{textcomp}
\usepackage{xcolor}

\usepackage[figuresleft]{rotating}
\usepackage{lineno}
\usepackage{booktabs}
\usepackage{algorithm}
\usepackage[noend]{algpseudocode}
\usepackage{tabu}
\usepackage[hang]{footmisc}
\usepackage[normalem]{ulem}
\usepackage{balance}
\usepackage{textcomp}
\usepackage{breqn}
\usepackage{tikz}

\def\BibTeX{{\rm B\kern-.05em{\sc i\kern-.025em b}\kern-.08em
    T\kern-.1667em\lower.7ex\hbox{E}\kern-.125emX}}

\title{Reinforcement Learning in Computing and Network
Convergence Orchestration}
\author{Mohan Wu $^1$, Boquan Cheng $^1$, Aidong Yang $^1$, Xiaozhou Ye $^1$, Ye Ouyang $^2$\\
1 Asiainfo Technologies (China), Inc., Beijing, China\\
2 Asiainfo Technologies (Guangzhou), Inc., Guangzhou, China}

\begin{document}

\maketitle
\begin{abstract}
    As computing power is becoming the core productivity of the digital economy era, the concept of Computing and Network Convergence (CNC), under which network and computing resources can be dynamically scheduled and allocated according to users' needs, has been proposed and attracted wide attention.
    Based on the tasks' properties, the network orchestration plane needs to flexibly deploy tasks to appropriate computing nodes and arrange paths to the computing nodes. This is a orchestration problem that involves resource scheduling and path arrangement. 
    Since CNC is relatively new, in this paper, we review some researches and applications on CNC. Then, we design a CNC orchestration method using reinforcement learning (RL), which is the first attempt, that can flexibly allocate and schedule computing resources and network resources. Which aims at high profit and low latency. Meanwhile, we use multi-factors to determine the optimization objective so that the orchestration strategy is optimized in terms of total performance from different aspects, such as cost, profit, latency and system overload in our experiment. The experiments shows that the proposed RL-based method  can achieve higher profit and lower latency than  the greedy method, random selection and balanced-resource method. We demonstrate RL is suitable for CNC orchestration. This paper enlightens the RL application on CNC orchestration.   \footnote{“This work has been submitted to the IEEE for possible publication. Copyright may be transferred without notice, after which this version may no longer be accessible.”}
\end{abstract}


\section{Introduction}
\hspace{1.5em}
The most critical resources in the digital economy era are data, computing power, and algorithm. 
As more sensors are deployed in our daily life, richer data with high dimension can be collected everyday. It is challenging to analyse the big data due to its complex characteristics. As the big data has rich features, the analysis should be performed on a hardware equipped with strong computing power.

Computing power has become the most dynamic and innovative new productivity in the current digital economy era. It is a key indicator to measure the digital economy and the core productivity of the digital economy era. 
The computing power in a narrow sense is the amount of information data that a device can process per second based on changes in its internal state. The core of it is Central Processing Unit (CPU), graphics processing units (GPU), Field Programmable Gate Array (FPGA), Application Specific Integrated Circuit (ASIC) and other computing chips, which are carried by computers, supercomputers, servers, cloud clusters and various intelligent terminals. 
According to Moore's law \cite{schaller1997moore} -- the number of transistors on a microchip doubles every two years, though the cost of computers is halved, the computing power doubles every two years.
In a broad sense, computing power is the new productive force in the digital economy era and a solid foundation to support the development of the digital economy.

The Computing and Network Convergence (CNC) integrates computing power into the network. Many general technologies, for instance artificial intelligence, machine learning, block-chain, can be applied individually or jointly to optimize the management and orchestration of computing power. For example, the CNC may use machine learning method to find and schedule an optimal end-to-end path connecting between the access node and the computing node. Eventually, anyone in anywhere can have strong computing power as far as the network access is available.

In the future holographic intelligent society, various sensing terminals will generate massive amounts of raw data, whose processing requires a lot of computing power, and users from anywhere desire strong computing power. The CNC can satisfy the above needs, as it can orchestrate the network resources such as computing power based on the network status, the quality of service (QoS) and quality of experience (QoE). 

To guarantee the QoS and QoE, it critical to have appropriate answers for the following two question. Based on the preset criteria (e.g. minimum latency, cost, jitter, packet loss rate, or maximum profit), which computing nodes should be used? And which paths that connect access nodes and computing nodes are optimal?

As tasks become complicated, the traditional scheduling strategy may not meet the QoS and QoE requirement anymore. Hence, the traditional orchestration may not be the most suitable method for CNC. New network orchestration that can efficiently collaborate among clouds, networks, and edges, and be able to schedule tasks to the optimal computing nodes has not appeared yet. 

In order to solve the above problems, this paper proposes an orchestration method based on reinforcement learning (RL). Our method uses Deep Q Network (DQN) as the strategy network and Deep Q-Learning to train the network. The DQN learns to choose the optimal computing node associated with the path from access node to the computing node based on the tasks' requirement.
When a task comes to orchestration plane where the DQN is deployed, the orchestration plane does not need to search through all paths because the DQN will output the best path. Our experiments show that the proposed method can generate the highest accumulated profit and the lowest average latency, by which assigning tasks to the most optimal computing nodes.

The main contribution of this paper includes:
\begin{itemize}
    \item propose a RL-based orchestration method on CNC. Which can generate the highest accumulated profit and the lowest average latency compared with other benchmark methods; 
    \item use multi-factors to determine the optimization objective so that the orchestration strategy is optimized in terms of total performance from different aspects;
    \item demonstrate the suitability of using RL-based orchestration, which enlightens other RL related applications on network orchestration.
\end{itemize}

This paper is organized as follows: Section \ref{sec2} reviews the current researches on network orchestration. The network system is introduced in section \ref{sec3}. Four orchestration methods, including our proposed RL method and the other two for benchmark, are introduced in section \ref{sec4}. Section \ref{sec5} details the experiment with concrete evidence indicating the proposed RL method is better than the benchmark methods. The conclusion as well as the challenges need to overcome in the future research are in section \ref{sec6}. 

\section{Related Work}\label{sec2}
\begin{figure}
    \centering
    \includegraphics[width=\linewidth]{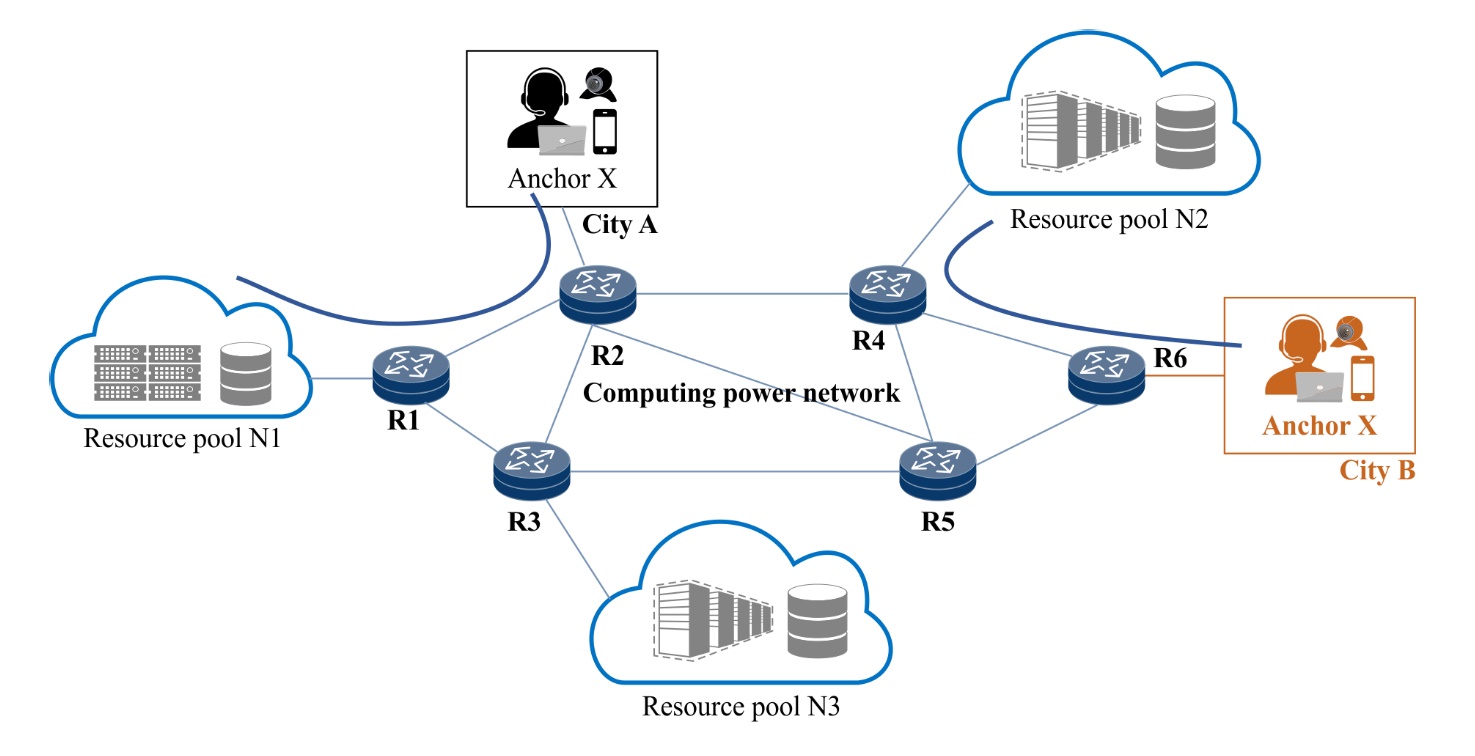}
    \caption{CNC in live broadcast scenario}
    \label{fig:full_system}
\end{figure}

\hspace{1.5em}
In September 2021, the International Telecommunication Union (ITU) published the standard of computing power network framework and architecture, which firstly attempted to standardize the structure of centralized computing power resources pools, see Fig. \ref{fig:full_system} as an example. 
Later on, ITU-T (ITU Telecommunication Standardization Sector) published the standard of Computing and Network Convergence (CNC) that unified the concepts of Computing Power Network (CPN), Computing-Aware Networking (CAN), and Computing Force Network (CFN).

With the advent of the 5G era, CNC has attracted more and more attention. Many researches have appeared to analyze and describe CNC from various aspects. Li et al. \cite{Cui2019FrameworkOC} first proposed a CFN framework that can send service requests to the best edge to improve overall system loading balance. This paper defines a routing protocol for computing resource information allocation and dynamic analysis based on late binding of control plane and data plane, respectively. Geng and Willis \cite{Geng2019ComputeFN} introduced some CNC application scenarios and proposed some CNC related requirements. He et al. \cite{HE2020Research} explained and analyzed the function and composition of each layer in the layered architecture of the CNC, and described the management of heterogeneous computing resources in the CNC. For quantification of computing power in the CNC, Li et al. \cite{Li2020Computing} normalized modeling the heterogeneous computing resources and proposed a grading standard for computing power. At the same time, they expounds the joint service capabilities of computing power, storage, and network to ensure business experience, and summarizes the service capability requirements of different types of businesses from a business perspective. Based on the TM Forum business process framework, Zhao et al. \cite{2020Operation} introduced specific operational functions. Additionally, these security concerns are addressed by adding security domains, and cryptography infrastructure is discussed in particular. Tian et al. \cite{2020An} gave an overview of CNC. Lyu and Liu \cite{2021Information} pointed out that the traditional computing power dominated by user services cannot meet the efficient needs of future collaborative computing, and tried to propose the research and implementation direction of distributed CNC.

In addition, more and more experiments related to CNC systems are emerging. K\'{r}ol et al. \cite{10.1145/3357150.3357395} design a computational graph representation for CNC framework with advantages of simplicity, performance, and failure resilience. Lei et al. \cite{2019Computing} proposed that CNC is a new multi-access edge computing, and presented a typical artificial intelligence application implementation system. This solution can effectively deal with the multi-level deployment of computing, storage, network and even algorithm resources in future business. Liu et al. \cite{0CATS} propose a multi-tier CNC and a computation offloading system with hybrid cloud and fog, define a weighted cost function consisting of delay, energy, and payment, and formulate a cost aware task scheduling (CATS) problem.

Recently, related articles on the arrangement and scheduling of CNC have also begun to appear. Huang et al. \cite{HUANG2021Computing} showed an IP network-based architecture of CNC is proposed.
The random selection, which randomly select which service to provision, and balanced-resource method, which selects the service that fits best the resources available, have been used for infrastructure resource orchestration \cite{natalino2018machine}. According to their experimental performance, these methods could be applied on CNC orchestration. 
However, the article on the algorithm of CNC orchestration has not yet appeared.



\section{System framework of the network}\label{sec3}
\subsection{The network system}
For the CNC framework, it has 2 classes of nodes -- network nodes and computing nodes. 
Different nodes have various transmission capacity (e.g. bandwidth) and computation resource (e.g. memory storage, CPU, GPU). 
However, the computation resource is limited, and the node transmits a limited number of tasks, which is within the transmission capacity, at the same time. 
If the working condition is over the transmission capacity or the computation resource limit, then the node is considered to be break down, leading to the failure of network system, in this paper. When the network is functional, we call the network is in health condition. 
It is worth noting that the performance of a node will drop significantly when its workload is very close to its limit. A good orchestration strategy should avoid scenarios that push any node to its limit. 

When a task arrives, it will get into the network from the access node (one kind of network node). Then, the orchestration plane will find computing node(s) and path(s) from access node to the computing node(s). The task will send to the destination for processing.
Each task will remain for several time slot according to the task requirement, and there could be multiple tasks arrives at the same time slot. Thus, how to make orchestration when the number of tasks in each time slot is large require careful calculation.

\subsection{Computing nodes}
There are various kinds of computing nodes.
A part of them, such as edge nodes, have low computation resource (weak computing power) and low latency (small transmission delay). 
On the contrary, another part of computation nodes, like cloud nodes, have rich computation resource (strong computing power) and high latency (long transmission delay). The remaining computation nodes are somewhere between the above two.
Furthermore, it is typical that the computation cost is much cheaper at the cloud nodes than that at the edge nodes. For example, the electricity cost of maintaining per unit of computing resource at a cloud center in the west of China is significantly lower than the electricity cost of maintaining per unit of computing resource at a edge in the east of China.

As a result, the advantages of cloud nodes are low cost and fast calculation speed, and its disadvantage is long transmission delay, whilst the strength of edge nodes is low low transmit delay, and its weaknesses are high cost and slow calculation speed.
Based on the characteristics of the task, some tasks (e.g. large tasks require intense computation but not sensitive to delay) are better to assign to cloud nodes, whilst others (e.g. small tasks are easy to compute but require low latency) are better to process at the edge nodes.

\subsection{Optimal orchestration}
\begin{figure}
    \centering
    \includegraphics[width=\linewidth]{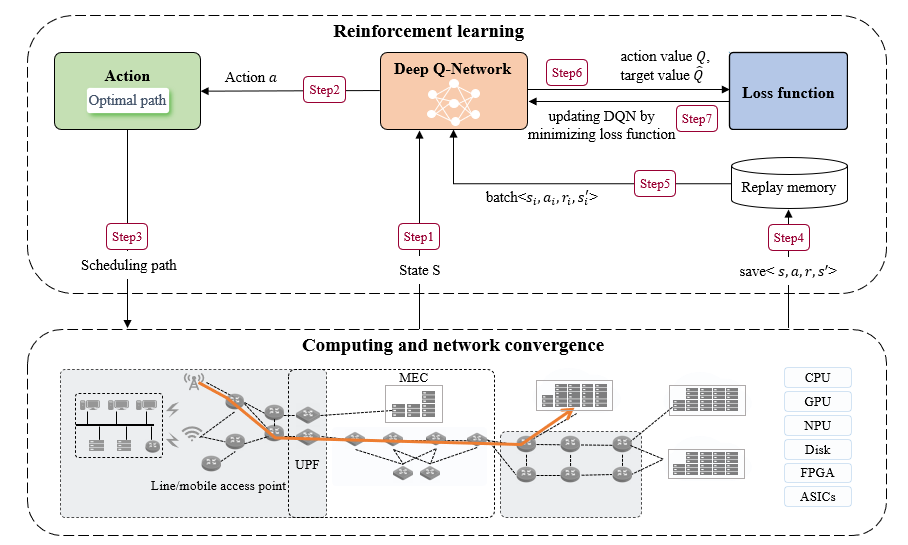}
    \caption{CNC orchestration by reinforcement learning.}
    \label{fig:my_label}
\end{figure}

Recall that CNC can collect information, including CPU usage, bandwidth, upload speed, download speed etc, for each network node and computation node. The orchestration efficiency can be enhanced, if these information are utilized properly during strategizing the computing node and the associated path from access node to the computing node. 
An orchestration strategy is to determine the optimal computing node as well as the path connecting between the access node and the computing node. As mentioned previously, different types of tasks have their most suitable computing nodes. Therefore, the orchestration strategy must take into account of the the task properties, including the required computation resource, the maximum latency and cost that can tolerate etc.
If the selected path meets the task's requirement, the task is accepted by the system. Otherwise, the task will be rejected by the system. 
Fig \ref{fig:my_label} demonstrates an example of CNC orchestration.

To evaluate the orchestration strategy, an objective function (criterion) is required. In particular, the optimal orchestration strategy should optimize the objective function, e.g. maximize the profit, minimize the cost and latency.  Given a set of tasks, the optimal orchestration should accept as many tasks as possible and schedule them to the suitable computing nodes so that the accumulated value of objective function (e.g. accumulated profit or accumulated cost) is optimized. Meanwhile, the orchestration strategy should avoid the network break down, which is a devastating destruction for the system. 
Moreover, it is better to prevent any node from reaching a level close to its workload limit. 

In the traditional setting, an amount of revenue will be given when accepting a task. 
Then, given a set of tasks, an intuitive objective function could be the total profit, which is equal to the sum of the revenues collected minus the sum of the system cost. To address the other factors like latency, one can add the factors as a penalized term to the objective function.  
The domain of this optimization problem includes the computing nodes and the paths from access nodes to the computing nodes. Since there are a large number of combinations of computing nodes and paths, the domain is large to explore, resulting a complex optimization problem.
We will use 4 methods to obtain the orchestration strategy, and evaluate them by the objective function as well as the factors including profit, latency, and decision time.

\section{Methods for optimal orchestration}\label{sec4}

\begin{table*}
\caption{Comparison between four methods}\label{comparision}
\begin{center}
\begin{tabular}{|l|l|l|p{2.3cm}|l|l|p{4cm}|}
\hline
Method & candidate paths & training & objective function & decision speed & profit & characteristics\\ \hline
random selection & required & no & no & fast & low & easy to deploy\\ \hline
balanced-resource & required & no & lowest workload & slow & medium & lowest breach chance, traditional method from expert experience\\ \hline
greedy method & required & no & highest task profit & slow & high & highest profit for each task\\ \hline
Reinforcement learning & no & required & high accumulated task profit& fast & high & do not need candidates, high accumulated profit \\ \hline
\end{tabular}
\end{center}
\end{table*}

The traditional methods, random selection, balanced-resource method \cite{natalino2018machine}, greedy method and the proposed RL method are introduced for the CNC orchestration to improve system performance. Their features are summarized in Table \ref{comparision}.

\subsection{Preliminary for the orchestration strategy assessment}
Suppose a sequence of tasks $\mathbf{M}=(M_1,\ldots,M_k)$ waiting for scheduling and orchestration. The current status of all nodes is observed, denoting the status as $\mathcal{S}$.
Given a task $M_i,\ i=1,\ldots,k$, one can search all computing nodes that can meet the task's requirement. Then, the candidate paths can be found, where candidate paths include all paths from the access node to the computing nodes that does not break down the system. 
Let $\mathrm{D}$ be the set of all paths in the network, and $\mathcal{D}_{M_i}$ be the set of candidate paths that can meet the  requirement of $M_i$. 
Then, the profit $p(M_i)$ for task $M_i$ is 
\begin{align}
    p(M_i)=R_{M_i} - C_{M_i}-(l^*_{M_i}-l_{M_i})^+, \, \text{for} \, d_i\in \mathcal{D}_{M_i},
\end{align}
where $R_{M_i}$ and $C_{M_i}$  are revenue and cost, $l^*_{M_i}$ and $l{M_i}$ are the actuarial and required latency respectively. Since reject a task cannot receive any revenue, $R_{M_i}=0$ if rejecting task $i$.
The cost and actuarial latency vary for different path $d_i$.
The $(\bullet)^+$ is a ReLU (Rectified Linear Unit) function, which is equal to itself only when its value is greater than 0, otherwise is equal to 0.

The orchestration optimization objective is 
\begin{align}
   \max \limits_{d_i\in \mathcal{D}_{M_i}}\sum_{i=1}^kp(M_i).
\end{align}

The remaining of this section introduces 3 methods to determine $d_i$.

\subsection{random selection}
The random selection includes the following two steps:
\begin{itemize}
    \item all candidate paths for task $M_i$ are searched, consisting the set  $\mathcal{D}_{M_i}$;
    \item use random number generator to sample a path $d_i$ from $\mathcal{D}_{M_i}$.
\end{itemize}

This two-step process keeps going for all tasks. Eventually, the orchestration strategy for the sequence of tasks are $d_1,\ldots,d_k$. The last node of $d_i$ indicates the selected computing node for task $M_i$. 

It is worth noting random selection does not maximize the profit (objective function), but it is easy for implementation. Also, it can guarantee the system never break down, since all paths are selected from the candidate paths and all nodes' workload are not over their limits (allow pushing the workload to the limit).
Therefore, random selection is a good benchmark for the experiment, which is also the benchmark  in \cite{natalino2018machine}.  



\subsection{balanced-resource}
The balanced-resource method includes the following three steps:
\begin{itemize}
    \item all candidate paths for task $M_i$ are searched, consisting the set  $\mathcal{D}_{M_i}$;
    \item check the workload of computing nodes in $\mathcal{D}_{M_i}$, and the computing node with the lowest workload will be chosen as the computing node for $M_i$;
    \item Once the computing node is fixed, find the best paths $d_i$ to the selected computing node in $\mathcal{D}_{M_i}$, where the best paths are in terms of the highest available bandwidth and lowest latency.
\end{itemize}

This three-step process keeps going for all tasks. Eventually, the orchestration strategy for the sequence of tasks are $d_1,\ldots,d_k$. 

By doing so, the workloads for computing nodes and network nodes are balanced, under which the network works most efficiently and safely according to experts' opinion. The system never break down by balanced-resource method since it selects the path with lowest workload.
However, this method does not maximize the profit (objective function). This method can be a benchmark from expert experience for the experiment, which is also a benchmark in \cite{natalino2018machine}.

\subsection{greedy method}
The balanced-resource method includes the following three steps:
\begin{itemize}
    \item all candidate paths for task $M_i$ are searched, consisting the set  $\mathcal{D}_{M_i}$;
    \item loop through all the paths in $\mathcal{D}_{M_i}$, find the best paths $d_i$, where the best paths are in terms of the highest profit for current task.
\end{itemize}

This two-step process keeps going for all tasks. Eventually, the orchestration strategy for the sequence of tasks are $d_1,\ldots,d_k$. 

By doing so, each task is assigned to the most profitable path. The system never break down by greedy method since it only choose path from candidate paths. Besides, this method is the intuitive method to maximize the profit (objective function), which is also a benchmark in \cite{greedy}.

\subsection{Reinforcement learning}
This is our proposed method for orchestration.
The reinforcement learning method is to train an agent (DQN) that can choose the optimal computing node and path. The input of the DQN is the task's ($M_i$'s) requirement, and the output of the DQN is the optimal path $d_i$. More theoretical details on RL are referred to \cite{sutton2018reinforcement}. 
To illustrate the RL method, we 
define state $S$, action $A$, reward $R$, and the agent using DQN.
\\\
\textbf{State $S$:} 
The state consists of all nodes information, including computing resource, current available bandwidth, and CPU processing speed, etc. The state also contains the task requirement like computation resource, bandwidth, latency, as well as task properties such as revenue. 
\\\
\textbf{Action $A$:}
The action space is the set of all paths $\mathcal{D}$. The agent can choose a path in $\mathcal{D}$ in each action.
\\\
\textbf{Reward $R$:}
It is intuitive to set 
\begin{align}
    R=\sum_{i=1}^kp(M_i)-\delta,
\end{align}
where $\sum_{i=1}^kp(M_i)$ is the accumulated profit in an episode, and $\delta$ is the penalty term for actions break down the system. When the network is in health condition, $\delta=0$.
According to our experiments, $\delta$ cannot be too large. If $\delta$ is too large, the agent likely learns rejecting all tasks for conservative behavior. 
\\\
\textbf{Agent:}
The agent makes the orchestration strategy. We use DQN to model the agent. Generally, one can apply more sophisticated RL methods, e.g. DDPG (Deep Deterministic Policy Gradient), TRPO (Trust Region Policy Optimization), PGB (Policy Gradient with Baseline).
Based on the current state $S$, the agent evaluates the best action $a$ to achieve the highest reward $R$. Before deploying the agent, it needs to be trained by past experience.

\section{Experimental assessment}\label{sec5}
\subsection{Experiment Setup}
To demonstrate the proposed method outperform the other two methods, we conduct simulation experiments based on a virtual network system. The system includes three types of computation nodes to represent the difference between mobile edge computing (MEC), edge cloud, cloud center with different resources capacity and positions in the network topology. We use a typical setup for the network system that has higher delay at nodes with higher loading and the latency between two nodes is determine by their distance. 

The tasks' properties are randomized within certain range for better generalization. For example, some tasks require a lot of computing resource with high latency tolerance, while some tasks require little computing resource but sensitive to latency. And the number of task in each time slot is set differently to test its affection.

It requires to train the agent before testing for RL, while random selection method, balanced-resource method and greedy method can apply directly on testing.
The experiment consists of a total of 20,000 tasks for RL training with 10 tasks per time slot and 1000 tasks for testing respectively.
Our experiment was conducted by using gym modulo to simulate the environment and using PyTorch to construct and train the agent in Python 3.8. 

\subsection{Accumulated Reward}
\begin{figure*} [t!]
	\centering
	\subfloat[\label{fig:3a}tasks reward with small number\\ tasks per time slot]{
		\includegraphics[width=0.4\linewidth]{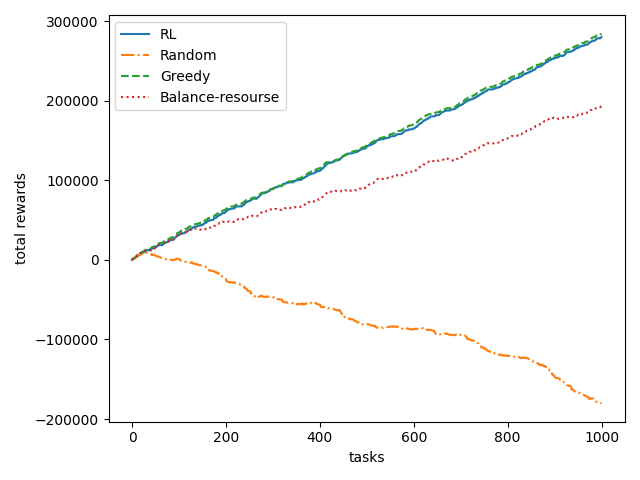}}
	\subfloat[\label{fig:3b}tasks reward with medium number\\ tasks per time slot]{
		\includegraphics[width=0.4\linewidth]{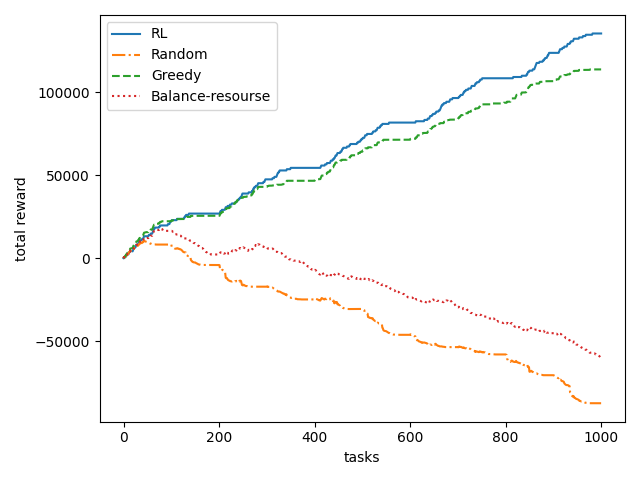}}
	\\
	\subfloat[\label{fig:3c}relation of total reward and tasks per time slot]{
		\includegraphics[width=0.4\linewidth]{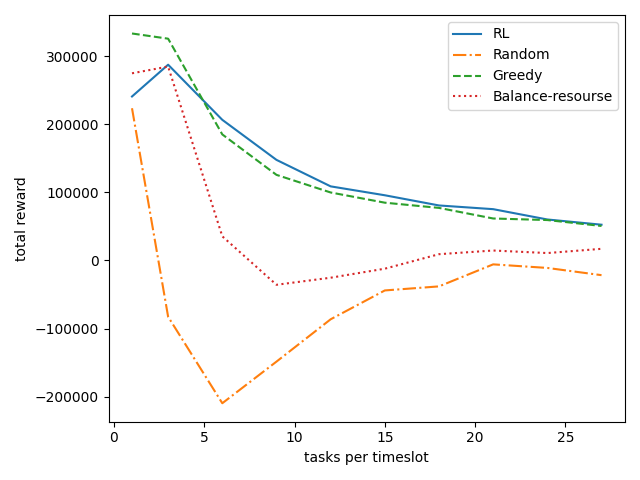} }
	\subfloat[\label{fig:3d}task latency with four methods]{
		\includegraphics[width=0.4\linewidth]{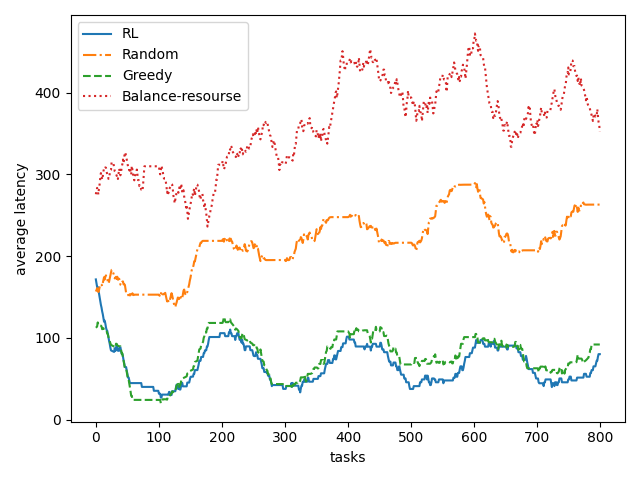}}
	\caption{(a) the accumulated task profit by RL, random selection, and balanced-resource method with 4 tasks in each time slot; (b) the accumulated reward for tasks by RL, random selection, and balanced-resource method with 10 tasks in each time slot; (c) the average latency of testing tasks by RL, random selection, and balanced-resource method; (d) the average decision time of testing tasks by RL, random selection, and balanced-resource method.}
	\label{fig:fig3} 
\end{figure*}

For random selection method, balanced-resource method and greedy method, the reward is the same as the profit since the available candidate paths will not break the system, and for RL method, the reward is profit with break down penalty $\delta$. We compare the accumulated reward with different task number in each time slot by the four methods in Fig \ref{fig:fig3} (a), (b) and (c). According to the figure (a), when the task number in each time slot is small, the greedy method generates highest accumulated reward; RL method is the second but very close to the greedy method; balanced-resource method is the third. Here the random selection method generate a negative accumulate rewards after 50 tasks, which is caused by huge latency when used resource is close to the limit. According to the figure (b), when the task number in each time slot is medium, the RL method generates highest accumulated rewards and greedy method is the second even though it achieves the best reward for every task. Here the behavior of random selection method and balanced-resource method are similar to each other, drop done to negative reward. Because when the task number increases, there is no enough resource for all the tasks, thus orchestrating more task into the system will cause the used resource close to the limit, leading to large latency. The reason why RL method is better is because RL agent learns the potential value of each task and reject some low reward task to provide resource to later high reward task. This shows the high accumulated reward advantage of RL is more significant as an increasing number of tasks for orchestration. 

To examine the effect of task number in each time slot, we test the performance of all four method with different tasks per time slot. As the figure (c) shows, when the task number is small (1-3), greedy method is the best, balanced-resource method and RL method are second and third respectively. This is when the resource is abundant and far from the resource limit. It is understandable that greedy method is the best orchestration method in this situation. When the task number per time slot is medium (4-24), RL method is the best. This is when the resource is not enough for all the task and used resource is close to the limit. When the task number per time slot is high (25-), RL method is still the best but perform very close to greedy method. This is where the resource is not enough for the tasks in even one time slot. The performance of all four method tend to be stable after more tasks per time slot, the reason is that the extra tasks will just be rejected due to no resources or low reward.  


\subsection{Latency and Decision Time}
In Fig \ref{fig:fig3} (c), we compare the average latency of 1000 tasks, which is an important factor for service-level agreement (SLA), by the 4 methods. The average latency by random selection is the highest and balanced-resource method is the second highest, whilst the average latency by greedy method and RL method is far less than the other two methods. The overall latency of RL method is lower than greedy method. Hence, the orchestration by RL is more efficient with lower latency than the other 3 methods. 

\begin{table}
\caption{Decision time between four methods}\label{decision time}
\begin{center}
\begin{tabular}{|l|l|l|l|l|l|}
\hline
Method & random & balanced-resource & greedy & RL\\ \hline
Decision time & 0.01s & 0.49s & 0.46s & 0.04s\\ \hline
\end{tabular}
\end{center}
\end{table}

The average of 1000 tasks' decision time, see table \ref{decision time}, that is the time required to obtain 1000 optimal path, is 0.04s by RL method. Compared with the 0.49 ms and 0.46 ms by balanced-resource method and greedy method, the decision time reduces 90\%, which shows orchestration by RL is more efficient than that by greedy method and balanced-resource method with respect to the time required for orchestration. The decision time of random selection is still the shortest, 0.01s. Recall that the network topology in our experiment is a simple version of that in real-world, the advantage of decision speed by RL is more meaningful in real network applications.

\section{Conclusion and challenges}\label{sec6}
\subsection{conclusion}
In this paper, we demonstrated the proposed RL method is applicable for CNC orchestration. 
Compared with the traditional method using the random selection, balanced-resource method and greedy method, the RL method improves the network efficiency with lower latency, lower decision time, and higher accumulated profit. The trained agent by RL can learn to select the optimal computing nodes for tasks based on the tasks' characteristics and reject tasks when the resources is full. The agent takes into account of reserving resources for future tasks, e.g. the resource is reserved for more profitable tasks in our experiment.  

Additionally, the trained agent is easy to deploy in orchestration control plane. It has great value to apply RL on the CNC orchestration.

\subsection{challenges in future research}
The RL method requires model training before deploying. It is challenging to train the agent when the network topology is complicated. Unfortunately, the network topology in real world is complicated. In order to over this challenge, some pre-processing, e.g. simplify the network topology by grouping network nodes based on experts' experience, may be needed.

The second challenge is how to mimic the network in real world as accurate as possible in the simulation environment. To fill the gap of simulation accuracy, it involves a lot of expert experience as well as advance technologies, such as machine learning methods, to refine the network simulator. 

\section*{Acknowledgment}
This work was supported by the Support Scheme of Guangzhou for Leading Talents in Innovation and Entrepreneurship (No: 2020010).

\bibliographystyle{IEEEtran}
\bibliography{IEEEabrv,reference}

\end{document}